\documentclass[preprints,article,accept,pdftex,moreauthors]{Definitions/mdpi} 

\firstpage{1} 
\pubvolume{1}
\issuenum{1}
\articlenumber{0}
\pubyear{2022}
\copyrightyear{2022}
\datereceived{} 
\dateaccepted{} 
\datepublished{} 
\hreflink{https://doi.org/} 
\pdfoutput=1

\Title{Tidal Deformability of Neutron Stars in Unimodular Gravity}

\TitleCitation{Title}

\Author{Rui-Xin Yang $^{1,\dagger}$, Fei Xie $^{1,\dagger}$ and Dao-Jun Liu  $^{1,}$* \orcidA{}}

\AuthorNames{Rui-xin Yang, Fei Xie and Dao-Jun Liu}

\AuthorCitation{Yang, R.; Xie, F.; Liu, D.-J.}

\address{%
Department of Physics and Center for Astrophysics, Shanghai Normal University, Shanghai 200234, China  
}

\corres{Correspondence: djliu@shnu.edu.cn}

\firstnote{These authors contributed equally to this work.}

\abstract{Unimodular gravity is a modified theory with respect to general relativity  by an extra condition that the determinant of the metric is fixed. Especially, if the energy-momentum tensor is not imposed to be conserved separately, a new geometric structure appears with potentially observational signatures. In this paper, we study tidal deformability of compact star in the unimodular gravity under the assumption of non-conserved energy-momentum tensor. Both the electric-type and magnetic-type quadrupole tidal Love numbers are calculated for neutron stars with polytrope model. It is found that the electric-type tidal Love numbers are monotonically increasing, but the magnetic-type ones are decreasing, with the increase of the non-conservation parameter. Compared with the observational data from detected gravitational-wave events, a small negative non-conservation parameter is favored.   }


\begin{document}


\section{Introduction}

 The tidal deformability, or equivalently the tidal Love number (TLN),  characterizes the degree of the response of a body to an external tidal force \cite{Poisson2014}. The observations of gravitational waves from coalescing neutron star and neutron-star-black-hole binaries can measure the tidal deformability of neutron stars \cite{Flanagan2008,Hinderer2010}, which enable us not only to reveal important information regarding their internal structure and the equation of state (EOS) of high-dense nuclear matter, but also to test the property of gravity in strong field regime \cite{Cardoso2017}. Therefore, it is interesting to quantify the effects from gravity on the TLNs in different gravitational theories. The theory of TLNs in general relativity (GR) has been developed elegantly \cite{Hinderer2008,Binnington2009,Damour2009}. The TLNs  of neutron stars, black holes and other compact objects in modified theories of gravity have been studied in a few cases in the past years (see, for example, Refs. \cite{Yagi2013,Yazadjiev2018,Silva2021}). 
In Ref.\cite{Meng2021}, Meng and Liu calculate  the TLNs of neutron stars in Rastall gravity and find that they are significantly smaller than those in GR. 

Unimodular gravity (UG) is another alternative formulation of the gravitational theory initiated by Einstein himself, shortly after the birth of GR \cite{Einstein1919}. The basic idea behind UG is that the determinant of the metric of spacetime is kept fixed instead of being a dynamical variable. This constraint reduces the symmetry of the diffeomorphism group to the group of unimodular general coordinate transformations. As a consequence,  the equations governing the dynamics of spacetime are the traceless Einstein equations. Thus, the vacuum energy in UG has no direct gravitational effect and the cosmological constant becomes simply an integration constant of the dynamics. Due to this property, Weinberg pointed out that UG could be used to deal with the cosmological constant problem \cite{weinberg1989}. 

The above form of UG works under an important hypothesis that the energy-momentum tensor is covariantly conserved.  It turns out that if the conservation of the energy-momentum tensor is imposed, at least in the classical level, UG does  correspond to GR with an additional integration constant associated to the cosmological term \cite{Abbassi2008} \footnote{Actually, even imposing the conservation of the energy-momentum tensor, some new features may appear at perturbative level, see Ref. \cite{Gao2014}.} (see, for a recent review, \cite{Carballo-Rubio:2022ofy}).  However, it should be kept in mind that the conservation of the energy-momentum tensor is not automatic in UG but is introduced as an additional assumption. If the energy-momentum tensor does not obey the usual conservation laws, the correspondence with GR is broken in the presence of matter.

UG with or without hypothesis of conservation of the energy-momentum tensor has been extensively investigated in the literature \cite{Finkelstein2001,Perez:2018wlo,Moraes2022,Nakayama2022,Almeida:2022qld}, particularly in the context of cosmology \cite{Shaposhnikov2009,Jain:2012gc,Garcia-Aspeitia:2019yni,Leon:2022kwn,Fabris2022} and in the quantization problem of gravity \cite{Unruh:1988in,Smolin2011,Yamashita2020}. Recently, the evolution of gravitational waves in UG without conservation of the energy-momentum tensor is considered and the  differernce with the usual signatures in GR is shown \cite{Fabris2022a}. In view of the extensive interest of UG in the community of gravitation, it is of importance to seriously investigate the stellar dynamics in UG.

The stellar dynamics in UG under the assumption of non-conserved energy-momentum is considered in \cite{AstorgaMoreno2019} and it is found that modifications due to the non-conservation of energy-momentum that lead to sizeable effects that could be constrained with observational data. Therefore, it is of interest to investigate the effect of the violation of the classical energy-momentum conservation on the tidal deformability of compact stars in UG. In the present paper, we want to calculate the tidal Love numbers of a neutron star in the fully relativistic polytrope model within the framework of UG.

The rest of the paper is organized as follows: In Sec.\ref{sec:UG}, we briefly review the theoretical framework of UG.
Sec. \ref{sec:NSinUG} is dedicated to the study of static, spherically symmetric solutions of the modified Tolman-Oppenheimer-Volkoff (TOV) equation under an assumption for the non-conservation of energy-momentum to account for stars described by a polytropic EOS. 
In Sec.\ref{sec:TLNs}, we study the static linearized perturbations of the neutron star and introduce the definition of TLNs. Then,  we perform a numerical analysis and explore the relationship between the TLNs and non-conservation parameter, obtaining modifications that could provide constraints on the non-conservation of energy-momentum. 
Finally,  in Sec. \ref{sec:discussion}, we present our conclusions and give some discussions. In the paper we work in geometric units $\left(c=G=1\right)$,  unless otherwise
noted.

\section{Unimodular gravity: action and field equations}\label{sec:UG}
As is pointed out in the introduction, UG is an alternative theory of gravity, which can be viewed as GR with a so-called unimodular condition \footnote{This constraint reduces the full diffeomorphism group of spacetime to transverse diffeomorphisms subgroup satisfying $g_{\mu\nu}\delta g^{\mu\nu}=0$.}
\begin{equation}\label{UC}
\sqrt{-g}=1.
\end{equation}
By introducing a Lagrange undetermined multiplier $\lambda$ into Einstein-Hilbert action, UG would be defined by the following action 
\begin{equation}
S=\frac{1}{16\pi}\int\left[\sqrt{-g}R+\lambda\left(\sqrt{-g}-1\right)\right]+S_{M},
\end{equation}
where $S_{M}$ denotes the action for some matter fields. The variation of $\lambda$ recovers the unimodular condition \eqref{UC}, and the variation with respect to $g^{\mu\nu}$ yields
\begin{equation}\label{EOM}
R_{\mu\nu}-\frac{1}{2}g_{\mu\nu}R-\frac{1}{2}g_{\mu\nu}\lambda=8\pi T_{\mu\nu},
\end{equation}
where
\begin{equation}
T_{\mu\nu} \equiv -\frac{2}{\sqrt{-g}}\frac{\delta S_{M}}{\delta g^{\mu\nu}}
\end{equation}
is the energy-momentum tensor of the matter field as usual. The trace of Eq.\,\eqref{EOM} gives the Lagrange multiplier, which reads
\begin{equation}
\lambda=-\frac{1}{2}\left(8\pi T+R\right).
\end{equation}
Substituting this result back into Eq.\,\eqref{EOM} leads to the trace-free version of Einstein field equations which will be used later
\begin{equation}\label{UG}
R_{\mu\nu}-\frac{1}{4}g_{\mu\nu}R=8\pi\left(T_{\mu\nu}-\frac{1}{4}g_{\mu\nu}T\right).
\end{equation}

If the covariant conservation law of the energy-momentum tensor $\nabla^{\mu}T_{\mu\nu}=0$ is assumed, then the Bianchi identities, i.e. the vanishing divergence of the Einstein tensor $\nabla^{\mu}G_{\mu\nu}=0$, imply that
\begin{equation}
8\pi T+R=4\varLambda,
\end{equation}
where $\varLambda$ is a constant of integration. Inserting it in Eq.\,\eqref{UG}, we obtain
\begin{equation}
G_{\mu\nu}+\varLambda g_{\mu\nu}=8\pi T_{\mu\nu},
\end{equation}
which means that UG is equivalent to GR with the cosmological constant appearing as an arbitrary integration constant, once the energy-momentum tensor conservation is imposed.

Thus, in order to find the deviation of stellar dynamics and tidal response from those in GR, we attempt to break the conservation of $T_{\mu\nu}$ in the following sections.

\section{Neutron Stars in UG}\label{sec:NSinUG}
To calculate the tidal Love numbers, we first need to get a neutron star solution in UG. To this end, let us  consider a static, spherically symmetric star solution, the metric of which in the spherical coordinates $\left\{t,r,\theta,\varphi\right\}$ can be written as
\begin{equation}\label{SSM}
g_{\mu\nu}^{\left(0\right)}dx^{\mu}dx^{\nu}=-e^{2\alpha\left(r\right)}dt^{2}+e^{2\beta\left(r\right)}dr^{2}+r^{2}d\theta^{2}+r^{2}\sin^{2}\theta d\varphi^{2},
\end{equation}
where both $\alpha$ and $\beta$ are  only functions of radial coordinate $r$. Obviously, the metric above does not satisfy the unimodular condition \eqref{UC}, however, we can perform a coordinate transformation
\begin{equation}
dr=\frac{e^{-\left[\alpha\left(r\right)+\beta\left(r\right)\right]}}{r^{2}}dy,\ x=\cos\theta,
\end{equation}
and rewrite the Eq.\,\eqref{SSM} in the unimodular coordinate system $\{t,y,x,\varphi\}$ as
\begin{equation}
g_{\mu\nu}^{\left(0\right)}dx^{\mu}dx^{\nu}=-e^{2\alpha\left(y\right)}dt^{2}+\frac{e^{-2\alpha\left(y\right)}}{r^{4}\left(y\right)}dy^{2}+\frac{r^{2}\left(y\right)}{1-x^{2}}dx^{2}+r^{2}\left(y\right)\left(1-x^{2}\right)d\varphi^{2},
\end{equation}
where the Eq.\,\eqref{UC} is satisfied explicitly. In fact, it can be proved that the physical results are the same in both systems \cite{AstorgaMoreno2019}. Therefore, we still work in the standard spherical coordinates.

Furthermore, the matter within the star takes an isotropic perfect fluid described by the energy-momentum tensor
\begin{equation}\label{EMT}
T_{\mu\nu}^{\left(0\right)}=\left[\rho\left(r\right)+p\left(r\right)\right]U_{\mu}U_{\nu}+p\left(r\right)g_{\mu\nu}^{\left(0\right)},
\end{equation}
where $\rho$ and $p$ are the energy density and pressure measured by a comoving observer, respectively, while $U_{\mu}=\left(-e^{\alpha},0,0,0\right)$ is the 4-velocity.

Outside the star, there is no matter. Then, from Eqs.\,\eqref{UG} and \eqref{SSM} it follows that the vacuum solution in UG has the Schwarzschild-de Sitter form
\begin{equation}
e^{2\alpha}=e^{-2\beta}=1+\frac{c_{1}}{r}+c_{2}r^{2}
\end{equation}
with two integration constants $c_{1}$ and $c_{2}$. Clearly, $c_2$ plays the role of cosmological constant, which can be neglected for our purpose. Therefore, the Schwarzschild solution in UG is obtained with $c_{1}=-2M$ and $M$ the mass of the star.

Inside the star, feeding back Eqs.\,\eqref{SSM} and \eqref{EMT} into UG field equations \eqref{UG}, we obtain that
\begin{align}
m^{\prime}&=\frac{m}{r}-\left(r-2m\right)f+4\pi r^{2}\left(\rho+p\right),\label{dmdr}\\
f^{\prime}&=\frac{4\pi r^{3}\left(\rho+p\right)\left(1+rf\right)-2m}{r^{2}\left(r-2m\right)}-2f^{2},\label{dfdr}
\end{align}
where the primes denote  derivatives with respect to the radial coordinate $r$.
Here, the mass function $m$ is  introduced by
\begin{equation}
e^{2\beta}=\left[1-\frac{2m\left(r\right)}{r}\right]^{-1},
\end{equation}
 and $f\equiv\alpha^{\prime}$.

Two significant facts should be illustrated at this point. First, it is known that, in GR, the stellar structure and spacetime geometry is built by three independent differential equations, in which the $\theta\theta$-component is equivalent to the conservation $\nabla^{\mu}T_{\mu r}=0$; however, here in UG, there are only two independent equations, namely Eqs.\,\eqref{dmdr} and \eqref{dfdr}, due to the constraints of the unimodular condition \eqref{UC} on the metric or the traceless property of the equations of motion \eqref{UG}, conservation of energy is not essential at least from a purely theoretical point of view. Second, unlike in GR, one cannot eliminate all the second derivatives both of $\alpha$ and $\beta$ from the field equations \eqref{UG} under the ansatz \eqref{SSM} and \eqref{EMT} by some algebraic manipulations, this is why we bring in a function $f$ here.

As mentioned above, the only two independent equations are not enough to determine the four dependent variables, $m$, $f$, $\rho$ and $p$. In order to close the system, following Ref.\cite{AstorgaMoreno2019}, we assume that the non-conservation behavior of energy-momentum tensor takes the following form
\begin{equation}\label{non-con}
\nabla^{\mu}T_{\mu\nu}=k\rho\delta^{r}_{\ \ \nu}
\end{equation}
where $k$ is a constant. When $k$ is set to be zero, we hope that the GR solutions could be recovered. In Ref.\cite{AstorgaMoreno2019}, Astorga-Moreno et al. showed that a positive $k$ may lead to larger star compactness than GR ones. After some straightforward computations, Eq.\,\eqref{non-con} becomes
\begin{equation}\label{TOV}
p^{\prime}=-\left(\rho+p\right)f+k\rho,
\end{equation}
which is the modified TOV equation, or the equation of hydrostatic equilibrium in UG, except that $f$ cannot be written explicitly. In addition, one need an EOS to supplement the system of equations, i.e. to specify a relation between density and pressure, $p=p\left(\rho\right)$. A simple but somewhat realistic  model describing the EOS for nuclear matter is known as polytrope model, which is given by
\begin{equation}\label{EOS}
p=\kappa\rho^{\Gamma},
\end{equation}
where $\kappa$ is a constant and the exponent $\Gamma$ is the so-called adiabatic index, associated with the polytropic index $n$ via $\Gamma=1+1/n$.

Now, we can solve this system numerically to get an equilibrium configuration of the star with Newtonian mass $M$, as long as a set of initial conditions at the center of the star are specified appropriately. To do so, let us consider the Taylor series expansion of the functions $m$, $\rho$ and $f$ near $r=0$, and fix the value of $\rho^{\prime\prime}\left(0\right)$ with $k=0$ to be the same as in the GR solution, such that the contribution of cosmological constant is eliminated. Next, just integrate Eqs.\,\eqref{dmdr}, \eqref{dfdr}, \eqref{TOV} and \eqref{EOS} outward in the domain $0<r<R$, where $R$ is the surface of the star determined by the condition $p=0$.

\section{Static Linearized perturbations and TLNs}
\label{sec:TLNs}
Suppose we have a neutron star immersed in an external tidal field which, for example, arise from its companion in a binary system. Consequently, the original spherical body is deformed by the tidal force, and develope some mass multipole moments in response to the tidal field. Tidal Love numbers characterize the deformability of the stellar objects, the bigger TLNs, the bigger deformation.

In short, the spacetime geometry is perturbed by the external tidal field and we can write
\begin{equation}\label{per}
g_{\mu\nu}=g_{\mu\nu}^{\left(0\right)}+h_{\mu\nu},
\end{equation}
where $g_{\mu\nu}^{\left(0\right)}$ is the background metric defined by Eq.\,\eqref{SSM} in the previous section, whereas $h_{\mu\nu}$ is a small perturbation owing to the tidal field, satisfying $|h_{\mu\nu}|\ll|g_{\mu\nu}^{\left(0\right)}|$. In Regge-Wheeler gauge, according to the parity of the spherical harmonics under the rotation on 2-sphere $S^{2}$, the perturbation metric $h_{\mu\nu}$ can be decomposed into even and odd parts according to parity under the rotation in $(\theta,\phi)$-plane \cite{Binnington2009,Cardoso2017}
\begin{equation}
h_{\mu\nu}=h_{\mu\nu}^{\text{even}}+h_{\mu\nu}^{\text{odd}},
\end{equation}
with
\begin{equation}\label{perE}
h_{\mu\nu}^{\text{even}}=
\begin{bmatrix}
-e^{2\alpha}H_{0}\left(r\right)&H_{1}\left(r\right)&0&0\\
H_{1}\left(r\right)&e^{2\beta}H_{2}\left(r\right)&0&0\\
0&0&r^{2}K\left(r\right)&0\\
0&0&0&r^{2}\sin^{2}\theta K\left(r\right)\\
\end{bmatrix}
Y^{lm}\left(\theta,\varphi\right),
\end{equation}

\begin{equation}\label{perO}
h_{\mu\nu}^{\text{odd}}=
\begin{bmatrix}
0&0&h_{0}\left(r\right)S^{lm}_{\theta}\left(\theta,\varphi\right)&h_{0}\left(r\right)S^{lm}_{\varphi}\left(\theta,\varphi\right)\\
0&0&h_{1}\left(r\right)S^{lm}_{\theta}\left(\theta,\varphi\right)&h_{1}\left(r\right)S^{lm}_{\varphi}\left(\theta,\varphi\right)\\
h_{0}\left(r\right)S^{lm}_{\theta}\left(\theta,\varphi\right)&h_{1}\left(r\right)S^{lm}_{\theta}\left(\theta,\varphi\right)&0&0\\
h_{0}\left(r\right)S^{lm}_{\varphi}\left(\theta,\varphi\right)&h_{1}\left(r\right)S^{lm}_{\varphi}\left(\theta,\varphi\right)&0&0\\
\end{bmatrix}
.
\end{equation}
Here $Y^{lm}$ is the scalar spherical harmonics, $S^{lm}_{\theta}\equiv-\partial_{\varphi}Y^{lm}/\sin\theta$ and $S^{lm}_{\varphi}\equiv\sin\theta\partial_{\theta}Y^{lm}$ are two axial vector spherical harmonics. All the functions in $h_{\mu\nu}$ are independent of time $t$ which means that the tidal field is assumed to be stationary for our purpose. Acctually, this is a typical scenario occuring in the inspiral stage of binary system.

Substituting \eqref{per} into the left hand side of Eq.\,\eqref{UG} and keeping only up to first order terms in $h_{\mu\nu}$, one reaches the linearized UG equations
\begin{equation}\label{perUG}
\delta R_{\mu}^{\ \ \nu}-\frac{1}{4}\delta_{\mu}^{\ \ \nu}\delta R=8\pi\left(\delta T_{\mu}^{\ \ \nu}-\frac{1}{4}\delta_{\mu}^{\ \ \nu}\delta T\right),
\end{equation}
where
\begin{equation}
\delta T_{\mu}^{\ \ \nu}=\text{diag}\left[-\delta\rho\left(r\right),\delta p\left(r\right),\delta p\left(r\right),\delta p\left(r\right)\right]Y^{lm}\left(\theta,\varphi\right)
\end{equation}
is the corresponding fluctuations of matter field to the background \eqref{EMT}.

Henceforth, we shall focus our attention on the lowest quadrupolar order $\left(l=2\right)$ in perturbations, because it dominates the tidal deformation of the stars. Because of the spherical symmetry of the system, we set the magnetic quantum number $m=0$ without loss of generality. For a non-rotating object, the even-parity sector decouples completely from the odd-parity sector in the linear level and vice versa, we therefore can discuss them individually, the former and the latter are related to the TLNs of electric-type and magnetic-type, respectively.

\subsection{Even-parity sector}
In this circumstance, the linearized equations \eqref{perUG} along with Eq.\,\eqref{perE} yields in turn: $H_{1}=0$, $H_{2}=-H_{0}$, $K^{\prime}=-H_{0}^{\prime}-2H_{0}\alpha^{\prime}$, and
\begin{equation}
K=\frac{1}{2}e^{-2\beta}\left[-r^{2}H_{0}^{\prime}\alpha^{\prime}+H_{0}\left(1-3e^{2\beta}+r\alpha^{\prime}+r\beta^{\prime}+2r^{2}\alpha^{\prime^{2}}\right)\right].
\end{equation}
Using these results, we finally obtain a single 2-order homogeneous differential equation for $H_{0}\equiv H$ as
\begin{equation}\label{H}
H^{\prime\prime}+\mathcal{P}\left(r\right)H^{\prime}+\mathcal{Q}\left(r\right)H=0,
\end{equation}
where the coefficients are
\begin{equation}
\mathcal{P}=\frac{-1+e^{2\beta}+r^{2}f^{\prime}+2r\left(1+2rf\right)\left[f-2\pi r\left(\rho+p\right)e^{2\beta}\right]}{r^{2}f}
\end{equation}
and
\begin{equation}
\mathcal{Q}=\frac{2\left[-1+2r^{2}f^{\prime}+2rf\left(1+rf\right)\right]}{r^{2}}-4e^{2\beta}\left[\pi\left(1+\frac{1}{c_{s}^{2}}\right)\frac{p^{\prime}}{f}+\frac{1+4\pi r^{3}\left(\rho+p\right)f}{r^{2}}\right],
\end{equation}
where $c_{s}^{2}=dp/d\rho$ is the speed of sound in the fluid.

Outside the star, we have $\rho=p=0$, $m=M$ and $f=M/r\left(r-2M\right)$, so Eq.\,\eqref{H} reduces to
\begin{equation}
H^{\prime\prime}+\frac{2\left(r-M\right)}{r\left(r-2M\right)}H^{\prime}-\frac{2\left(2M^{2}-6Mr+3r^{2}\right)}{r^{2}\left(r-2M\right)^{2}}H=0.
\end{equation}
The general solution to above equation is found to be
\begin{equation}\label{Hsol}
H=a_{1}P_{2}^{(2)}\left(\frac{r}{M}-1\right)+a_{2}Q_{2}^{(2)}\left(\frac{r}{M}-1\right),
\end{equation}
where functions $P_2^{(2)}$ and $Q_2^{(2)}$ are two independent associated Legendre functions with degree $2$ and order $2$. 
Here, two combination coefficients $a_{1}$ and $a_{2}$ are evaluated by integrating  Eq.\,\eqref{H} inside the star and matching to the exterior solution at the surface $R$.

\begin{figure}[ht]
    \includegraphics[width=0.8\columnwidth]{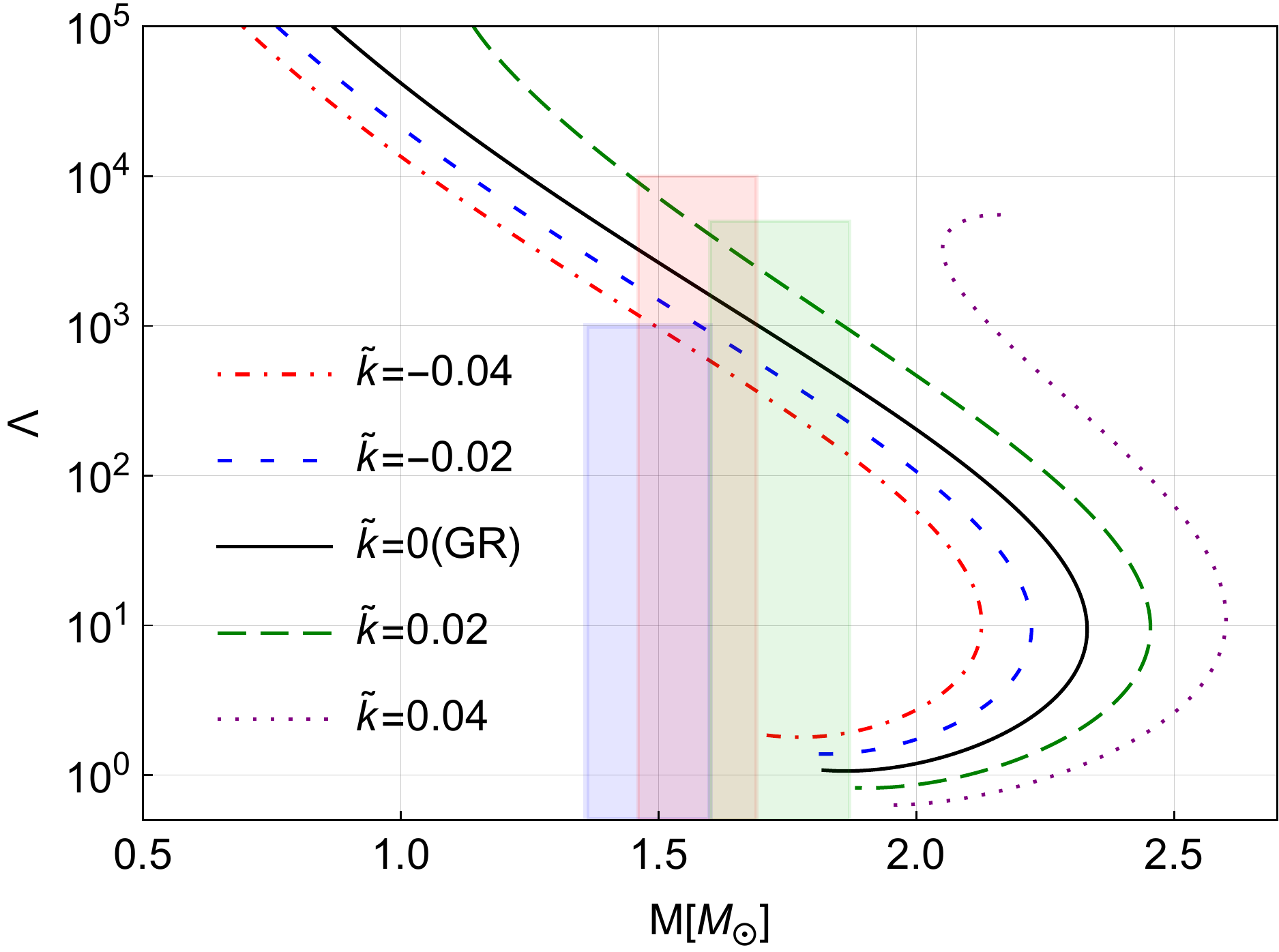}
    \centering
    \caption{The relationship between dimensionless tidal deformability and the mass of a star with adiabatic index $\Gamma =2$  for different values of dimensionless non-conservation parameter $\tilde{k}$. As a comparison, the constraints from gravitational wave observations  on the star are presented.  The blue shaded box is drawn based on the  $90\%$ confidence upper bound and  the $90\%$ confidence interval on the masses of the observed neutron stars from observations given by GW170817. Meanwhile, the green and red boxes represent the same bounds for the primary and secondary components with the low-spin priors in GW190425, respectively.
    }\label{fig:k2E}
   \end{figure}

Eventually, the asymptotic behaviour of the solution \eqref{Hsol} at infinity $\left(r\rightarrow\infty\right)$ determines the quadrupolar electric-type TLNs, it comes out
\begin{equation}
\begin{aligned}
k_{2}^{\text{E}}\equiv \frac{4}{15}C^{5}\frac{a_{2}}{a_{1}}=&\frac{8}{5}C^{5}\left(1-2C\right)^{2}\left[2C\left(y-1\right)+2-y\right]\left\{2C\left[4C^{4}\left(y+1\right)\right.\right.\\
&\left.+2C^{3}\left(3y-2\right)+2C^{2}\left(13-11y\right)+3C\left(5y-8\right)+6-3y\right]\\
&\left.+3\left(1-2C\right)^{2}\left[2C\left(y-1\right)+2-y\right]\ln (1-2 C)\right\}^{-1},
\end{aligned}
\end{equation}
where $C\equiv M/R$ stands for the compactness of the star, and $y\equiv RH^{\prime}\left(R\right)/H\left(R\right)$. For a comparison with observations, it is convenient to introduce the dimensionless tidal deformability 
\begin{equation}
    \Lambda \equiv \frac{2k_{2}^{\text{E}}}{3C^{5}}.
\end{equation}
In Figure \ref{fig:k2E}, we plot the dimensionless tidal deformability $\Lambda$ of a star made from the interacting Fermi gas (corresponding to the adiabatic index $\Gamma =2$) as a function of the star mass $M$ for different value of dimensionless non-conservation parameter $\tilde{k}=\left(1.73\times10^{6}\text{cm}\right)k$, where $k$ is defined in Eq.\,(\ref{non-con}) and the numerical factor is typical length scales for the interacting Fermi gas system. Furthermore, we compare them with the $90\%$ confidence upper bounds on $\Lambda$ for GW170817 \cite{Gw170817} (blue box) and GW190425 \cite{GW190425} (green and red boxes for the primary and secondary components) and the $90\%$ confidence intervals on the masses of the observed neutron stars \cite{Collier:2022cpr}. Clearly, both the mass and tidal deformability of the star are heavily influenced by the non-conservation of energy-momentum tensor.  The star mass and the tidal deformability are both monotonically increasing with the increase of the value of parameter $k$. Compared with the observations of gravitational wave, a negative $k$ is favored.    
\subsection{Odd-parity sector}
For odd-parity perturbations \eqref{perO}, from Eq.\,\eqref{perUG} we find $h_{1}=0$, and $h_{0}\equiv h$ satisfying the following equation
\begin{equation}
h^{\prime\prime}-4\pi r\left(\rho+p\right)e^{2\beta}h^{\prime}-2\left[f^{\prime}+2f^{2}-4\pi rf\left(\rho+p\right)e^{2\beta}+\frac{3e^{2\beta}}{r^{2}}\right]h=0.
\end{equation}
Outside the star, the above equation reduces to  be
\begin{equation}
h^{\prime\prime}+\frac{2\left(2M-3r\right)}{r^{2}\left(r-2M\right)}h=0,
\end{equation}
and the general solution of this simple equation can be expressed as 
\begin{equation}
h=b_{1}\left(\frac{r}{2M}\right)^{2}{ }_{2}F_{1}\left(-1,4,4;\frac{r}{2M}\right)+b_{2}\left(\frac{2M}{r}\right)^{2}{ }_{2}F_{1}\left(1,4,6;\frac{2M}{r}\right),
\end{equation}
where $b_{1}$ and $b_{2}$ are two undetermined constants, and $_{2}F_{1}$ denotes the hypergeometric function. Similarly, the quadrupolar magnetic-type TLNs reads now
\begin{equation}
\begin{aligned}
k_{2}^{\text{M}}\equiv -\frac{32}{3}C^{5}\frac{b_{2}}{b_{1}}=&\frac{8}{5}C^{5}\left[2C\left(z-2\right)-z+3\right]\left\{2C\left[2C^{3}\left(z+1\right)\right.\right.\\
&\left.+2C^{2}z+3C\left(z-1\right)-3z+9\right]\\
&+3\left[2C\left(z-2\right)-z+3\right]\ln\left(1-2C\right)\Big\}^{-1},
\end{aligned}
\end{equation}
with $z\equiv Rh^{\prime}\left(R\right)/h\left(R\right)$.

\begin{figure}[ht]
    \centering
    \includegraphics[width=0.8\columnwidth]{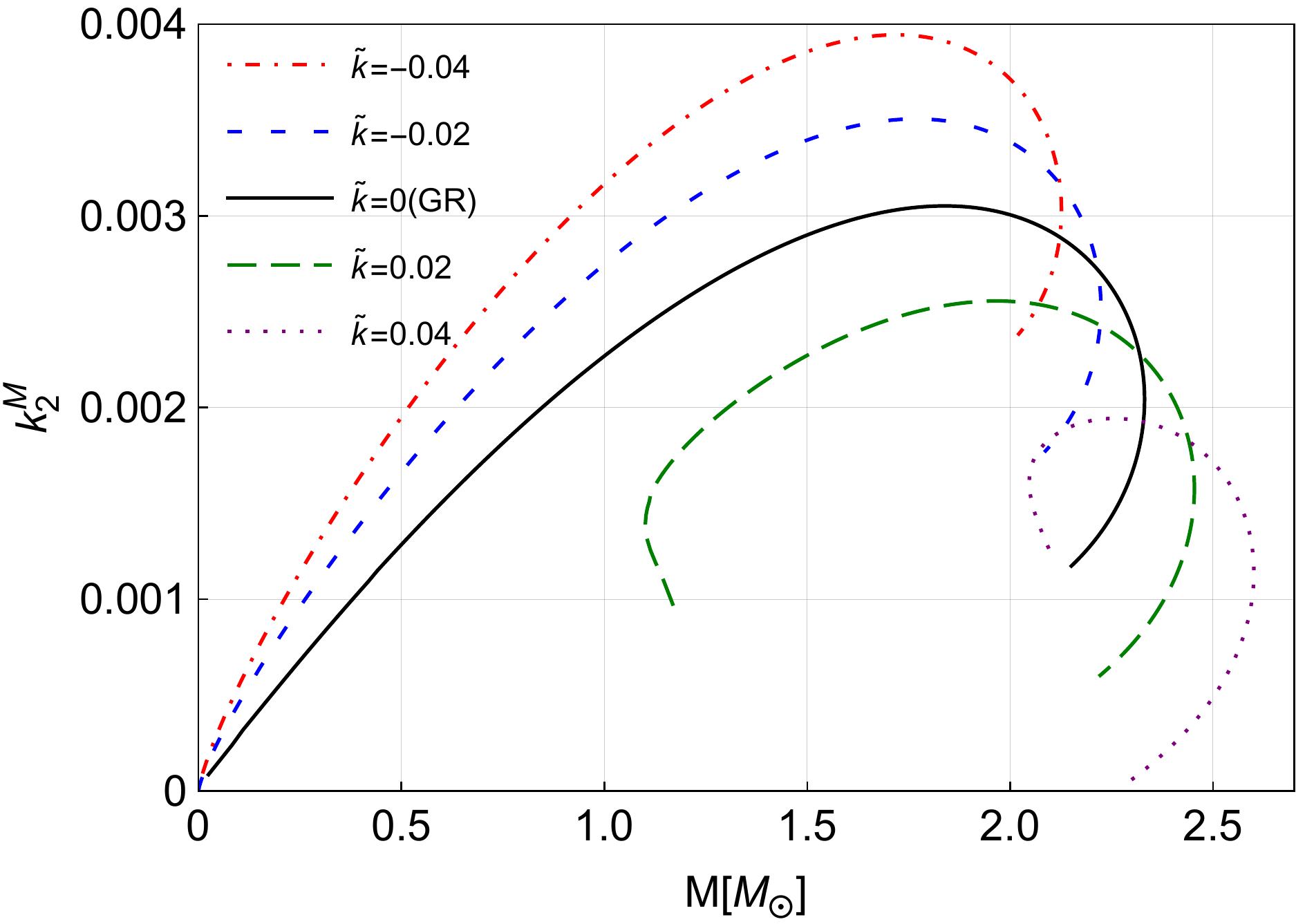}
    \caption{The magnetic-type quadrupolar TLNs $k_2^{\mathrm{M}}$ as a function of the mass of the neutron star $M$ for different values of the parameter $\tilde{k}$, where we choose  the adiabatic index $\Gamma =2$.  }
    \label{fig:k2M}
   \end{figure}

In Figure~\ref{fig:k2M}, the magnetic-type quadrupolar TLNs $k_2^M$ are plotted as a function of the star mass $M$ for the cases with different values of the parameter $\tilde{k}$. The EOS of the matter is chosen to be the same as in Figure~\ref{fig:k2E}. It is shown that for the same mass $M$, $k_2^M$ is monotonically decreased with the increase of the value of parameter $k$. 

\section{Conclusions and discussions}
\label{sec:discussion}

In this work, we have computed the TLNs of a spherically symmetric neutron star in the fully relativistic polytrope model within the framework of UG with non-conserved energy-momentum tensor. We show that non-conservation of the energy-momentum tensor has remarkable influence on the tidal deformability of the neutron star. For the case with the positive covariant divergence of the energy-momentum tensor, the electric-type TLNs of the star are increased relative to those in GR, whereas the magnetic-type ones are decreased. On the contrary, the tidal deformabilities of the star become  smaller than those in GR for the same star mass, when the energy-momentum tensor of matter has negative covariant divergence. Furthermore, we compare our results with the observational data from detected gravitatonal-wave events GW170817 and GW190425 and find that a negative non-conservation parameter, which indicates a negative covariant divergence of energy-momentum tensor of matter, seems to be more favored.

It is well-known that different EOS can lead to significant differences in macroscopic quantities of neutron stars, such as mass, radii, deformability, etc. Therefore, it is worthwhile to investigate the tidal deformability of neutron star with more realistic EOS (see, e.g., \cite{Friedman1981,Douchin:2001sv,PhysRevD.79.124032}) within the framework of UG. Our conclusions drawn above may have to be changed, due to the degeneracy between the modification of gravity and the EOS of nuclear matter in neutron stars. 

One the other hand, our results provide new evidence of the degeneracy, just as in Ref.~\cite{Meng2021}, where another kind of the non-conservation of energy-momemtum tensor are introduced. Note that the electric-type and magnetic-type TLNs show exactly the opposite behavior (that is, if the electric-type TLNs are depressed, the magnetic-type TLNs are raised) in the present work, however, in \cite{Meng2021} both of the two types of TLNs are depressed. Given that the electric-type TLNs are favored to be lowered for the star in the polytrope model by the current gravitational wave observations, it is expected that the next generation detectors are able to distinguish the different behaviors of the magnetic-type TLNs  shown in this work and Ref.~\cite{Meng2021}.

Finally, it should be pointed out that we have only assumed a special kind of violation of conservation of energy-momentum tensor as Eq.(\ref{non-con}) in this paper, it would be interesting to investigate the tidal deformabilities of the neutron stars with other forms of non-conservation of energy-momentum tensor within the framework of UG. In particular, the Newtonian limit of Eq.\,\eqref{non-con} does exist some anomalies. It implies that some screening mechanisms is essential as a supplement to the modified theory \cite{Vainshtein:1972sx,PhysRevD.69.044026,Brax2013}, such that the effect caused by Eq.\,\eqref{non-con} can be suppressed in weak gravitational fields, and pass stringent solar system tests of gravity. We leave these issues for future work.

\acknowledgments{ The authors are indebted to Xing-Hua Jin for helpful discussions. This work is supported by innovation programme of Shanghai Normal University  under Grant No. KF202147.}

\begin{adjustwidth}{-\extralength}{0cm}
\reftitle{References}
\bibliography{TLN_UG}
\end{adjustwidth}
\end{document}